\documentclass{iopconfser}
\usepackage{orcidlink}
\usepackage{amsmath,amssymb}
\usepackage[numbers,sort&compress]{natbib}
\usepackage{graphicx, subfigure}

\usepackage{acronym}
\acrodef{EMRI}[EMRI]{extreme mass-ratio inspiral}
\acrodef{CO}[CO]{compact object}
\acrodef{LISA}[\textit{LISA}]{\textit{Laser Interferometer Space Antenna}}
\acrodef{GW}[GW]{gravitational wave}
\acrodef{SNR}[SNR]{signal-to-noise ratio}
\acrodef{MBH}[MBH]{massive black hole}
\acrodef{MLP}[MLP]{\emph{multi-layer perceptron}}
\acrodef{CI}[CI]{credible interval}

\usepackage[normalem]{ulem} %% For striketrhough command

\usepackage{etoolbox}
\newtoggle{checklength}
\toggletrue{checklength} % As for publication
\begin{document}

\title{Constraints on the extreme mass-ratio inspiral population from LISA data}

\author{S~Singh$^{1}$\orcidlink{0000-0003-4881-1067}, C~E~A Chapman-Bird$^{2}$\orcidlink{0000-0002-2728-9612}, C~P~L~Berry$^{1}$\orcidlink{0000-0003-3870-7215} and J~Veitch$^{1}$\orcidlink{0000-0002-6508-0713}}

\iftoggle{checklength}{
\affil{$^1$Institute for Gravitational Research, University of Glasgow, Kelvin Building, University Ave., Glasgow, G12 8QQ, United Kingdom} 
\affil{$^2$Institute for Gravitational Wave Astronomy \& School of Physics and Astronomy, University of Birmingham, Edgbaston, Birmingham B15 2TT, UK}
}

\email{s.singh.3@research.gla.ac.uk}

%% 3 pages start here
\begin{abstract}
Gravitational waves from \aclp{EMRI} (\acsp{EMRI}), the inspirals of stellar-mass compact objects into massive black holes, are predicted to be observed by the \acl{LISA} (\acs{LISA}).
A sufficiently large number of \acs{EMRI} observations will provide unique insights into the massive black hole population.
We have developed a hierarchical Bayesian inference framework capable of constraining the parameters of the \acs{EMRI} population, accounting for selection biases.
We leverage the capacity of a feed-forward neural network as an emulator, enabling detectability calculations of $\sim 10^5$ \acsp{EMRI} in a fraction of a second, speeding up the likelihood evaluation by $\gtrsim6$ orders of magnitude.
We validate our framework on a phenomenological \acs{EMRI} population model.
This framework enables studies of how well we can constrain \acs{EMRI} population parameters, such as the slope of both the massive and stellar-mass black hole mass spectra and the branching fractions of different formation channels, allowing further investigation into the evolution of massive black holes.
\end{abstract}

\section{The promise of \acl{EMRI} observations}
\label{sec:detecting_EMRI}

Observations reveal a correlation between the properties of \acp{MBH} and their host galaxies~\cite{Richstone:1998ky, Ferrarese:2000se}. 
Despite their ubiquity, the origins and growth of \acp{MBH} remain poorly understood~\cite{Volonteri:2021sfo, Alexander:2025rtn}.
Within a few parsecs around these \acp{MBH}, dense nuclear stellar clusters host \acp{CO} such as stellar-mass black holes, white dwarfs, and neutron stars.
When a \ac{CO} forms a highly asymmetric-mass binary with a \ac{MBH} and over time spirals inwards, losing angular momentum and emitting \acp{GW}, we have an \ac{EMRI}~\cite{Amaro-Seoane:2007osp, Berry:2019wgg}.
The \ac{EMRI} spends $\sim10^5$--$10^6$ orbital cycles in the millihertz regime ($\sim10^{-4}$--$10^{-1}~\mathrm{Hz}$), where the \ac{LISA} is most sensitive~\cite{LISA:2024hlh}, before the \ac{CO} plunges into the \ac{MBH}.
With these long-lasting \ac{GW} signals from \acp{EMRI}, it is possible to constrain \ac{MBH} ($m_1$) and \ac{CO} ($m_2$) masses, \ac{MBH} spin ($a$), and orbital eccentricity ($e_0$, which we define at the start of observation) within $10^{-3}\%$ accuracy~\cite{Barack:2003fp,Babak:2017tow, Chapman-Bird:2025xtd}.
The precision of \ac{EMRI} measurements offers a unique opportunity to explore \ac{MBH} population \cite{Babak:2017tow, Gair:2008bx}.

% The precision of \ac{EMRI} measurements offers a unique opportunity to explore the environment around \acp{MBH} \cite{Babak:2017tow, Speri:2022upm}
% \CCB{explore the properties of MBHs and their environments?}.

In dense nuclear stellar clusters, several mechanisms have been proposed that could lead to \ac{EMRI} formation~\cite{Amaro-Seoane:2012lgq}.
One such pathway is the loss-cone channel, where the trajectory of the \ac{CO} is altered to a high eccentricity, low angular-momentum orbit, such that with each pass to the periapsis, orbital decay happens via \ac{GW} emission~\cite{Lightman:1977zz, Merritt:2013awa}.
% \CCB{imprecise wording as it's not the CO that is altered but its trajectory. Maybe ``where two-body relaxation processes place the \ac{CO} on a high eccentricity, ...''}  
To sustain this process, the nuclear stellar cluster must maintain a sufficiently high \ac{CO} density by refilling the loss cone continually, a condition facilitated by \emph{weak}~\cite{1977ApJ...216..883B} or \emph{strong}~\cite{Alexander:2008tq} mass segregation.
The \ac{MBH} environment may also influence \ac{EMRI} formation.
The presence (\emph{wet} \acp{EMRI}~\cite{Pan:2021ksp, Lyu:2024gnk, Pan:2021oob}) or absence (\emph{dry} \acp{EMRI}~\cite{Alexander:2017aln, Cui:2025bgu}) of gases affects the eccentricity and inclination of the \ac{CO}, thereby leaving an imprint on the orbital parameters and the rate of \acp{EMRI}.
% Gases can influence the orbital evolution by inducing torques on the orbiting \ac{CO}, effectively enhancing \ac{EMRI} formation~\cite{Pan:2021oob}.
Given the variety and complexity of the astrophysical processes influencing \ac{EMRI} formation and the absence of direct observational evidence, these astrophysical processes remain poorly constrained, and the relative contribution of each formation channel to the total \ac{EMRI} rate remains uncertain~\cite{LISA:2022yao}.

By combining multiple \ac{EMRI} observations from \ac{LISA} data, it should be possible to constrain both the astrophysical processes governing \ac{EMRI} formation and their relative contributions to the overall population~\cite{Gair:2010yu, Chapman-Bird:2022tvu, Langen:2024ygz}.
To perform inference at the population level, we need to perform hierarchical inference.
This framework accounts for uncertainties in individual \ac{EMRI} parameter estimates~\cite{Mandel:2018mve, Chapman-Bird:2022tvu}, enabling population-level inference unbiased by selection effects.
% \CCB{Maybe join sentences: ``...level, we require a hierarchical inference framework that accounts for uncertainties in...''}
% Hierarchical inference combines information from multiple \ac{EMRI} observations by linking the parameters inferred for individual events to a common population model~\cite{Mandel:2018mve, Chapman-Bird:2022tvu}.
% This approach explicitly incorporates measurement uncertainties from each event while constraining the hyperparameters that describe the underlying astrophysical population

% While doing the hierarchical inference, we must include selection effects.
Since only sources above a detection threshold will be observed, we must correct for undetected events in the population inference. 
We consider using an \ac{SNR}-based selection function.
% Agnostic to any particular strategy for EMRI detection, we model the detectability $P_\mathrm{det}(\boldsymbol{\theta})$ of EMRI signals in terms of their SNRs.
This function quantifies the probability of detecting an event from a given population model having the parameters $\boldsymbol{\Lambda}$.
It is defined as $\alpha(\boldsymbol{\Lambda}) = \int \mathrm{d}\boldsymbol{\theta} \, {P}_{\mathrm{det}}(\boldsymbol{\theta})p_\mathrm{pop}(\boldsymbol{\theta}|\boldsymbol{\Lambda})$, where $\boldsymbol{\theta}$ represents \ac{EMRI} parameters (e.g., component masses and spins), $p_{\mathrm{pop}}$ is the population distribution (e.g., a power law for the \ac{MBH} mass spectrum), and ${P_\mathrm{det}}$ is the detection probability for a given set of $\boldsymbol{\theta}$~\cite{Mandel:2018mve}.
%we will use parametric models for .. This line does not make sense because we are talking about the problem of the selection function and not how we have improved in our model.
This integral $\alpha(\boldsymbol{\Lambda})$ defines the fraction of the population that would be detectable.
Accurately modelling this selection function is crucial to ensure unbiased inference of the underlying astrophysical population \cite{Mandel:2018mve}.

%  %  THIS NEEDS DISCUSSION
% ==========================================
%The selection function can be estimated as a Monte Carlo integral, requiring samples to be drawn from a population model and the \ac{SNR} computed for each.
%This introduces a two-fold problem: first, an accurate Monte Carlo sum requires at least $\sim10^5$--$10^6$ draws from the population distribution and secondly, calculating \acp{SNR} for all the drawn events. 
Evaluating the \acp{SNR} across parameter space to calculate $P_\mathrm{det}$ is expensive, even when using graphics processing unit acceleration for waveforms~\cite{Chapman-Bird:2025xtd}. 
This makes evaluating $\alpha$ expensive even for one point in the population parameter space. 
For a full population inference, we must explore many values of $\boldsymbol{\Lambda}$ and hence recalculate $\alpha$ many times.  
%Even with the acceleration provided by an \ac{SNR} emulator, performing this many evaluations still takes several seconds on GPU hardware.
As a result, the direct use of these calculations in hierarchical population inference is computationally prohibitive without additional acceleration or approximation strategies.
To overcome this, we introduce machine learning approaches to speed up the calculation of selection effects.

\section{Leveraging machine learning techniques for population inference}
\label{sec:population_inference}

We use Monte Carlo integration to calculate $\alpha$. 
To overcome the computational bottleneck of drawing $\sim10^5$--$10^6$ samples and evaluating the \ac{SNR} for each, we extend the \texttt{poplar} machine learning framework~\cite{Chapman-Bird:2022tvu} for fast \ac{SNR} and selection-function estimation.
We deploy two \ac{MLP} neural networks~\cite[Section 8.2.1]{Acquaviva:2023gkb} to emulate (i) the \ac{SNR} and (ii) the selection function.
Our framework incorporates several advancements over the previous \texttt{poplar} version~\cite{Chapman-Bird:2022tvu}, including more realistic \ac{EMRI} scenarios such as: random plunges within the finite \ac{LISA} observation window, as well as populations informed by astrophysical processes governing \ac{MBH} masses, spins and orbital configurations for both gas-rich~\cite{Pan:2021ksp, Lyu:2024gnk} and gas-poor~\cite{Alexander:2017aln, Cui:2025bgu} environments. 
Together, these extensions enable efficient and more physically grounded population inference across a much broader and more realistic parameter space.

\acp{MLP} are a class of fully connected neural networks where each neuron in a layer links to all in the next layer.
They are effectively capture non-linear relationships in high-dimensional mappings~\cite[Chapter 6]{Goodfellow-et-al-2016}, and are thus well-suited to emulating the \ac{EMRI} \ac{SNR} $\rho(\boldsymbol{\theta})$ and the selection function $\alpha(\boldsymbol{\Lambda})$.
The number of neurons and layers in the \acp{MLP} can be scaled according to problem complexity, making them an effective surrogate for the computationally expensive \ac{EMRI} \ac{SNR} and selection function evaluation.

\begin{figure*}
    \centering
    \includegraphics[width=7.8cm,height=7.8cm]{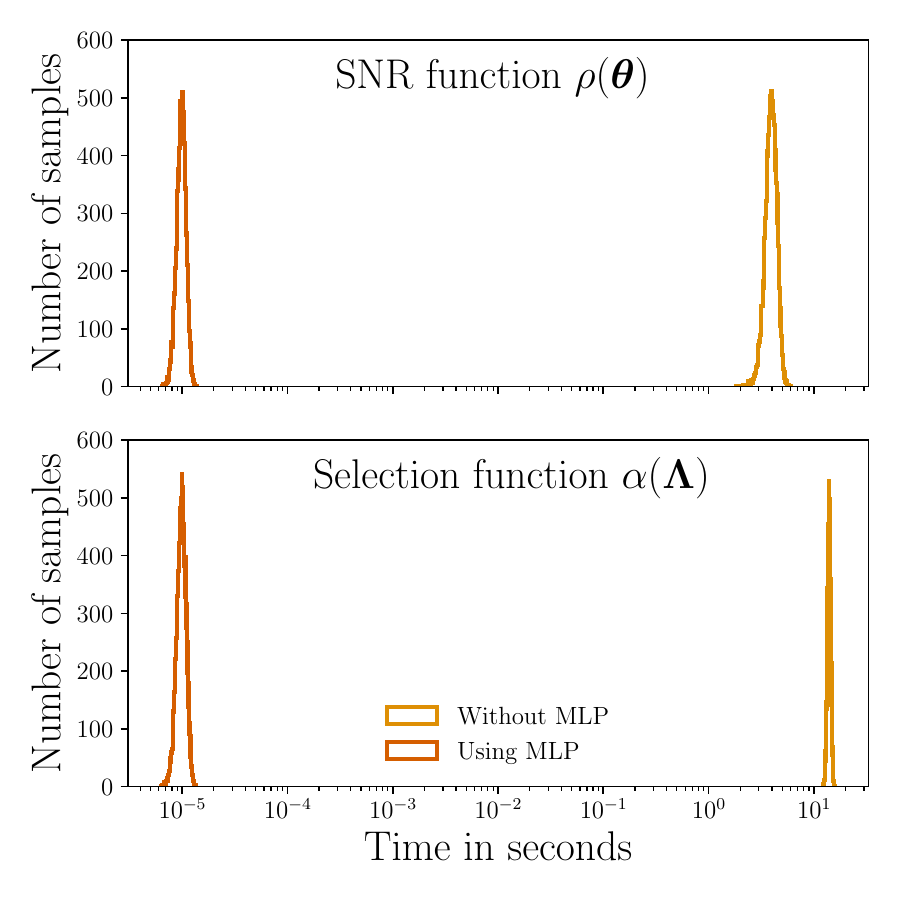}
    \includegraphics[width=7.8cm,height=7.8cm]{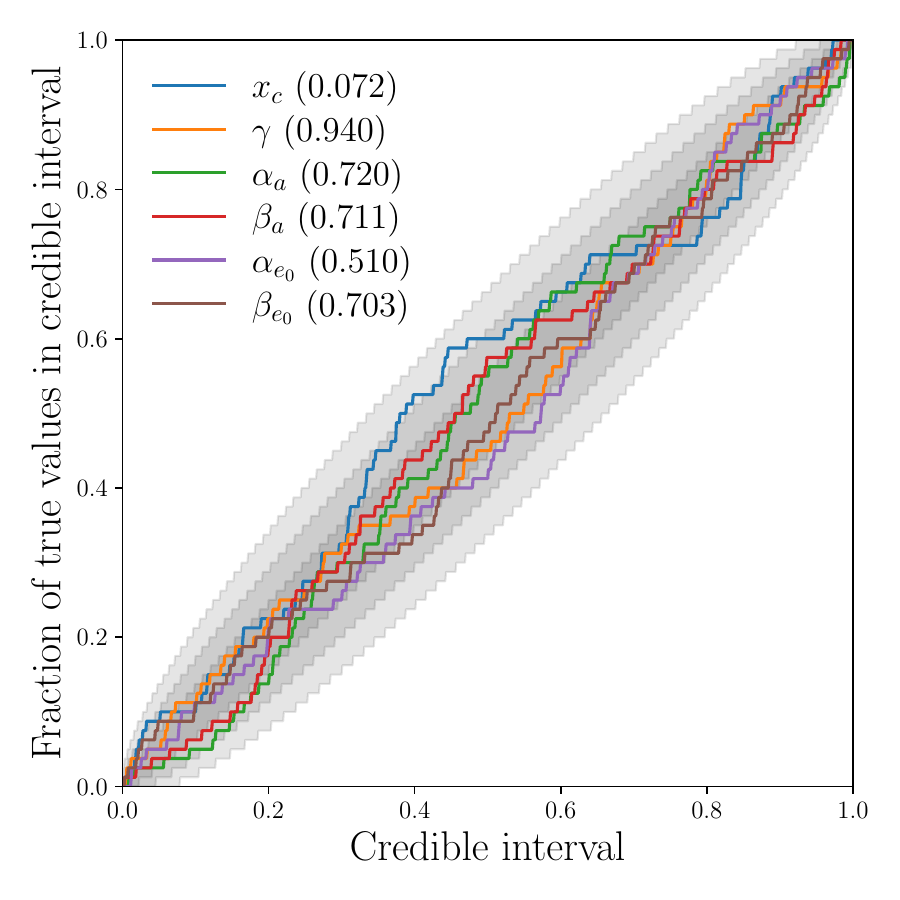}
    \caption{\emph{Left:} The time in seconds for \ac{SNR} (\emph{top}) and selection function (\emph{bottom}) evaluations with and without using \acp{MLP}.
    % For the selection function case without \acp{MLP}, the \ac{SNR} is computed using the \ac{SNR}-\ac{MLP}.
    \emph{Right:} Results from $100$ simulated populations sampled from the prior in Table~\ref{tab:population_parameters}. Grey bands indicate cumulative $68\%$, $95\%$, and $99.7\%$ intervals (in fading opacity). 
    Coloured lines show fractions of simulated values for each $\boldsymbol{\Lambda}$ parameter. 
    The corresponding $p$-values, comparing the distribution to the expected diagonal, are given in parentheses in the legend.}
    \label{plot_results}
\end{figure*}

We first train an \ac{MLP} to emulate the \ac{SNR} function, tuning both network complexity (neurons and layers) and other hyperparameters (learning rate, batch size and iteration count) based on validation loss to ensure stable convergence.
Training uses the AdamW optimiser~\cite{Loshchilov:2017bsp} with a decaying learning rate, and the data are min--max rescaled to normalise \ac{SNR} values.
To reflect observational conditions, we simulate \acp{EMRI} \acp{SNR} over a 4-year \ac{LISA} mission period, assuming continuous observation, and allowing for some \acp{CO} to plunge after observations end: plunge times are sampled uniformly across times from 1 to 6 years from the start of the mission.
The emulation results in a speed-up of $\sim10^5$ times compared to direct \ac{SNR} computation (Fig.~\ref{plot_results}a), enabling efficient evaluation of the detection probability.

% \textbf{Selection function} :
With the \ac{SNR} model trained, we address the second computational challenge, which is evaluating the selection function via Monte Carlo sum, which is both time-consuming and prone to noise unless a large number of samples is used.
To resolve this, we train another \ac{MLP} to map from the astrophysical parameters {($\boldsymbol{\Lambda}$)} of the population model to the selection function. 
We kept our model configuration similar to the \ac{SNR} model and applied a min--max rescaling to the selection function values.
This reduces the time for calculating the selection function, giving a speed-up of $\sim10^6$ times (Fig.~\ref{plot_results}a).

\begin{table}
    \centering
    \begin{tabular}{c c c}\hline
        \ac{EMRI} parameter $\theta_i$ & Functional form $p_{\mathrm{pop}}(\theta_i|\boldsymbol{\Lambda})$ & Population parameters $\Lambda_k$ \\\hline\hline
        Log \ac{MBH} mass $\xi = \log_{10}(m_1/M_\odot)$ & $C_\xi{x_c}^{-1} \left(\xi/{x_c} \right)^{\lambda}\exp\left( - \xi\lambda/x_c \right)$ & $x_c$ \\
        Log \ac{CO} mass $\zeta = \log_{10}(m_2/M_\odot)$ & $C_\zeta\exp({-\zeta^2/2})[1+\mathrm{erf}(\gamma\zeta/\sqrt{2})]$ 
        & $\gamma$ \\
        \ac{MBH} spin magnitude $a$ & $C_aa^{\alpha_a - 1}(1 - a)^{\beta_a - 1}$ & $\alpha_a, \beta_a$ \\
        Eccentricity $e_0$ & $C_{e_0}e_0^{\alpha_{e_0} - 1}(1 - e_0)^{\beta_{e_0} - 1}$ & $\alpha_{e_0}, \beta_{e_0}$ \\\hline
    \end{tabular}
    \caption{Distributions and functional forms used to model the population. 
    The distribution for each $\theta_i \in\boldsymbol{\theta}$ is assumed independent of the others. 
    We set the parameter $\lambda$ to a fiducial value of $\lambda = 7$.
    The normalisation constants $C_i$ depend on the corresponding population parameters.}
    \label{tab:population_parameters}
\end{table}

%\section{Results and conclusion}
To verify that our selection-function emulation is a sufficiently accurate approximation for hierarchical inference, we perform a large suite of simulated population analyses.
We used our emulation in the hierarchical inference framework for a phenomenological population described in Table~\ref{tab:population_parameters}.
The \ac{MBH} mass spectrum follows a Schechter distribution~\cite{1976ApJ...203..297S}; spin~\cite{Berti:2008af} and eccentricity~\cite{Mancieri:2025cmx} reflect typical high- or low-spin (eccentric) populations, while the \ac{CO} distribution reflects mass-segregation regimes depending on the abundance of high-mass \acp{CO}~\cite{Alexander:2008tq}.
In Fig.~\ref{plot_results}b, we show the results aggregated over many independent simulations, each using different injected population parameters drawn from the prior in Table~\ref{tab:population_parameters}.
The results from these independent simulations are shown in Fig.~\ref{plot_results}b, as a probability--probability (P--P) plot.
The P--P test is a statistical tool used to validate the consistency of parameter estimation~\cite{Cook:2006zir}.
It compares the \acp{CI} from an ensemble of simulated analyses.
Under a well-calibrated inference framework, the injected parameters should fall within the $q\%$ \acp{CI} for $q\%$ of analyses.
The P--P plot is a cumulative distribution of the \ac{CI} quantiles, where ideal inference follows the diagonal.
For our case of multivariate posteriors, P--P tests are applied to marginal distributions, and $p$-values are then combined, assuming independent population parameters~\cite{2017arXiv170706897H}, to assess overall parameter-estimation consistency \cite{Chapman-Bird:2022tvu, Romero-Shaw:2020owr}.
We obtained a combined $p$-value of $0.72$, which is consistent with the posteriors being well calibrated, with the lowest $p$-value of $0.07$ for the population parameter $x_c$.
Our results demonstrate that the emulation can efficiently account for the biases in the population inference, confirming its suitability for accurate population-level analyses.

\section{Summary}

We have demonstrated the use of \acp{MLP} to construct a computationally efficient, unbiased hierarchical inference framework that accounts for selection effects.
% in the analysis of \ac{EMRI} populations. 
By emulating both the \ac{SNR} and the selection function using \acp{MLP}, we avoid the bottlenecks associated with \ac{SNR} evaluation and large-scale Monte Carlo sum.
% Our \ac{SNR} emulator, trained on a realistic 4-year \ac{LISA} observation window, including scenarios with and without \ac{CO} plunge, achieves a speed-up of $\sim10^5$ times, which is currently the state-of-the-art emulation for \ac{EMRI}-\acp{SNR}.
Our \ac{SNR} emulator, trained on a realistic 4-year \ac{LISA} observation window, including scenarios with and without \ac{CO} plunge, achieves a speed-up of $\sim10^5$ times, representing a significant advancement in \ac{EMRI}-\ac{SNR} emulation capabilities.
Similarly, the selection function model accelerates evaluation by $\sim10^6$ times.
This allows performing hierarchical inference for a more complicated \ac{EMRI} population.
% encapsulating various astrophysical processes and their contribution to the total population.
% This allows performing hierarchical inference for a more complicated \ac{EMRI} population, which can encapsulate various astrophysical processes and their contribution to the total population. 
% The emulation tools we have developed thus represent a crucial step toward enabling realistic, high-dimensional population analyses with future \ac{LISA} data.

%% 3 pages to here
\iftoggle{checklength}{
\section*{Acknowledgments}
SS acknowledges support from the University of Glasgow.
CEAC-B acknowledges past support from STFC studentship 2446638, and current support from UKSA grant UKRI971.
This work was supported in part by STFC grant ST/V005634/1.
}

\bibliographystyle{iopart-num}
\bibliography{emri-refs}

\end{document}